\documentclass[aps,prb,amsmath,amssymb,showpacs,showkeys,twocolumn,groupedaddress]{revtex4}
\usepackage{dcolumn}
\usepackage{bm}
\usepackage[pdftex]{graphicx}

\begin{document}

\title{Microcanonical Monte Carlo approach for computing melting curves by atomistic simulations}

\author{Sergio Davis}
\email{sdavis@cchen.cl}

\affiliation{Comisión Chilena de Energía Nuclear, Casilla 188-D, Santiago, Chile}

\author{Gonzalo Guti\'errez}
\email{gonzalo@macul.ciencias.uchile.cl}

\affiliation{Grupo de Nanomateriales, Departamento de F\'{i}sica, Facultad de
Ciencias, Universidad de Chile, Casilla 653, Santiago, Chile}

\date{\today}

\begin{abstract}
We report microcanonical Monte Carlo simulations of melting and superheating of
a generic, Lennard-Jones system starting from the crystalline phase. The isochoric 
curve, the melting temperature $T_m$ and the critical superheating temperature $T_{LS}$ 
obtained are in close agreement (well within the microcanonical temperature
fluctuations) with standard molecular dynamics one-phase and two-phase methods. These 
results validate the use of microcanonical Monte Carlo to compute melting points, 
a method which has the advantage of only requiring the configurational degrees of freedom. 
Our findings show that the strict preservation of the Hamiltonian dynamics does not
constitute a necessary condition to produce a realistic estimate of $T_{LS}$ and
the melting point, which brings new insight on the nature of the melting transition.
These results widen the use and applicability of the recently developed Z method
for the determination of the melting points of materials.
\end{abstract}



\maketitle

\section{Introduction}

Melting curves of materials at extreme conditions are fundamental pieces of knowledge in 
the fields of materials science~\cite{Alfe2004}, geology~\cite{Belonoshko2000}, planetary
sciences~\cite{Oganov2005, Cavazzoni1999}, mechanical engineering~\cite{Padture2002}, 
condensed matter physics~\cite{Datchi2000}, among others, not to mention the
renewed interest in the melting mechanisms from the point of view of fundamental
science~\cite{Tallon89, Jin2001, Forsblom2005}. In both areas computer simulations play an 
increasingly important role, and development of new methods for computing
melting points, together with further improvement of existing methods, is a
crucial piece for future progress in the field. The current techniques used for the determination 
of melting curves via atomistic computer simulation (either from first-principles calculations or 
using semi-empirical interatomic potentials) can be divided into two categories: 
\emph{coexistence} simulations and \emph{one-phase} simulations.

The usual approach to simulating coexistence is the two-phase~\cite{Belonoshko1994,Morris1994} 
method. In this method a mixed sample, composed of different phases (in the case of melting, solid and liquid), is
simulated, and the thermodynamic conditions for coexistence of the two phases
are explored. For instance, in the microcanonical ensemble, the two-phase method
proceeds by choosing a total energy for which coexistence is observed and computing the average 
temperature, which is then associated with the melting temperature $T_m$. In the variants of 
the two-phase method where temperature is controlled (i.e., where simulations are carried out in the canonical or
isothermal-isobaric ensemble), an initially guessed temperature interval containing $T_m$ is
narrowed down systematically in order to constrain $T_m$ between an upper bound $T_0$ and 
a lower bound $T_1$, $T_0$ being such that it leads to an homogeneous solid phase and $T_1$ 
to an homogeneous liquid phase.
     
Regarding the one-phase methods, thermodynamic integration can be  
used~\cite{Sugino1995, DeWijs1998, DeKoning1999, Donadio2010} at constant volume to 
compute the Helmholtz free energy differences $\Delta F$ between the solid and liquid phase as a
function of temperature, and therefore to obtain the melting point the temperature $T_m$ 
for which $\Delta F(T_m)=0$. The same procedure can be applied in the
isothermal-isobaric ensemble to find the melting point at constant pressure $P$, by means of 
equating the Gibbs free energies of the different phases ($\Delta G(P, T_m)=0$) or even in the 
microcanonical ensemble by equating their entropies, $\Delta S(E_m)=0$ where $E_m$ is 
the internal energy of melting.

Most of these methods are implemented in ensembles different from the microcanonical, 
due to the simplicity of fixing the temperature or pressure as control
parameters to exactly their desired values. There are cases, however, where averages under 
such ensembles differ considerably from the (in principle) exact microcanonical averages. 
Ensemble equivalence is, in most cases, guaranteed in the thermodynamic limit (although
there are examples of systems where does not hold~\cite{Gulminelli2002,Campa2009}), and therefore, 
for small enough systems the appropriate course is to compute microcanonical
averages. It is mostly because of these limitations that the microcanonical approach~\cite{Pearson1985, 
Lustig1998, Gross2000, Gross2005} has regained interest when computing thermophysical properties 
and in the study of phase transitions for finite-size systems, such as metallic clusters~\cite{Westergren2003} 
and proteins~\cite{Junghans2006, HernandezRojas2008, Bereau2011}.

It is well known that early one-phase simulations of melting (such as the somewhat na\"{\i}ve idea of just heating the
solid using velocity rescaling until melting is observed) attempted before the
use of two-phase simulations suffer the phenomenon of \emph{superheating}, that
is, the melting temperature is overestimated. A relatively recent approach to 
determine the melting point using atomistic computer simulations is the Z method~\cite{Belonoshko2006}, 
which is a microcanonical one-phase method taking into account (and in fact, based on) the
superheating effect. \emph{Empirically} it has been found that, when starting from the ideal 
crystalline structure and increasing the total energy at fixed volume $V$, there is a well defined 
maximum (for the solid phase), $E_S(T_{LS}; V)$ where $T_{LS}$ corresponds to the limit of superheating. 
Increasing the energy beyond $E_S$ by a small amount $\delta E$, the solid spontaneously melts at
$E_S+\delta E \approx E_S$, but due to the increase in potential energy, namely
the latent heat of fusion, temperature decreases. The interesting fact is that the final 
temperature after melting at $E_S$ seems to coincide with the melting point $T_m$ obtained from other
methods. Thus the following equivalence is established in practice (so far
without clear theoretical foundations, but nevertheless supported by ample
evidence from numerical simulations),

\begin{equation}
E_S(T_{LS}; V) = E_L(T_m; V) \equiv E_{LS}.
\end{equation}

The procedure for the Z method computation of the melting point is then as
follows: at a fixed volume, the ($E$, $\big<T\big>$) points from different simulations draw a ``Z'' shape 
(hence the name of the method). In this Z-shaped curve the sharp inflection at the higher 
temperature corresponds to $T_{LS}$ and the one at the lower temperature to
$T_m$. Thus, knowledge of the lower inflection point for different densities
allows the determination of the melting curve for a particular range of pressures.

The Z method achieves the same precision in the determination of $T_m$ as the two-phase
method, but using only half the atoms (only a single phase is simulated at any

It has been proposed that this success of the Z method relies on sampling from
``genuine'' Hamiltonian dynamics, without resorting to fictitious forces such as
the ones arising from thermostat algorithms. In fact, the Z method has indeed
been tried under the Nos\'e-Hoover thermostat, leading to a lower value of
$T_{LS}$ relative to the microcanonical MD implementation. Moreover, $T_{LS}$
seems to be connected to anomalous diffusion time scales, as recently suggested~\cite{Davis2011}, 
which leads to the following question: \emph{to which extent is $T_{LS}$ a
dynamical phenomenon, and therefore dependent on the Hamiltonian dynamics?} Could the 
same results obtained in MD be also obtained following a stochastic dynamics, such 
as the one generated by microcanonical Monte Carlo (MC) methods? From the point of view of a
basic understanding of the nature of $T_{LS}$ it should be useful to clarify its dependence on 
strictly following the deterministic Hamiltonian trajectories.

In a practical sense, if following those trajectories is not required for a reliable
determination of the isochoric curve, it would widen the spectrum of possible
methods for computing melting points. If we could disregard the momentum degrees
of freedom it would be possible to afford larger system sizes, which is critical
in first-principles atomistic simulations.

In this paper we attempt to answer these questions. We present results indicating than 
a fully stochastic implementation of the Z method is possible, being in close agreement 
with the standard molecular dynamics implementation.

The paper is organized as follows. First, the MC formulation for
the microcanonical ensemble used in this work is presented, followed by the simulation 
details. Next, we describe the results obtained from comparison of MC and MD
simulations. Finally we summarize our findings.

\section{Microcanonical Monte Carlo}

We will consider a classical system of $6N$ degrees of freedom (3N momenta, denoted
collectively by $\mathbf{p}$, 3N coordinates denoted by $\mathbf{r}$), with
Hamiltonian

\begin{equation}
\mathcal{H} = \frac{\mathbf{p}^2}{2m} + \Phi(\mathbf{r}).
\end{equation}

The probability of the system having phase space coordinates $(\mathbf{r},
\mathbf{p})$ at total energy $E$ is given by 

\begin{equation}
P(\mathbf{r}, \mathbf{p}; E) = \frac{1}{\Omega(E)}
\delta(E-\mathcal{H}(\mathbf{r},\mathbf{p})),
\label{ProbDelta}
\end{equation}

where

\begin{equation}
\Omega(E) = \int d\mathbf{r} d\mathbf{p} \delta(E-\mathcal{H}(\mathbf{r},\mathbf{p}))
\end{equation}
is the density of states having energy $E$. Given that the dependence of the
Hamiltonian on $\mathbf{p}$ is fully known, those degrees of freedom can be
integrated out explicitly~\cite{Severin1978,Pearson1985}. To do this, we separate 
$\mathcal{H}$ inside the delta function and use 

\begin{equation}
\int d\mathbf{p} \delta(E-\mathbf{p}^2/2m-\Phi(\mathbf{r}))
\rightarrow \int_{\Sigma_p} \frac{d\Sigma_p}{|\nabla (\mathbf{p}^2/2m)|}
\end{equation}
where the last integral is over the $(3N-1)$-dimensional surface $\Sigma_p$
defined by $$|\mathbf{p}|=\sqrt{2m(E-\Phi(\mathbf{r}))}.$$ After this we can
rewrite the probability in Eq. \ref{ProbDelta} as

\begin{equation}
P(\mathbf{r}; E) = \frac{1}{\Omega(E)}
\Theta(E-\Phi(\mathbf{r}))\sqrt{E-\Phi(\mathbf{r})}^{3N-2},
\label{ProbRay}
\end{equation}
where now the density of states $\Omega(E)$ can be written as 
\begin{equation}
\Omega(E) = \int d\mathbf{r} \Theta(E-\Phi(\mathbf{r}))\sqrt{E-\Phi(\mathbf{r})}^{3N-2},
\end{equation}
and $\Theta$ is Heaviside's step function.

Equation \ref{ProbRay} leads to the following Metropolis acceptance probability~\cite{Ray1991},

\begin{equation}
P(\mathbf{r}_1 \rightarrow \mathbf{r}_2) = \min\left(1,
\sqrt{\frac{E-\Phi(\mathbf{r_2})}{E-\Phi(\mathbf{r_1})}}^{3N-2}\right).
\end{equation}

This rule makes it possible to simulate a system in the microcanonical ensemble
without incorporating the momentum degrees of freedom explicitly. It also
avoids the use of a ``demon'' to impose conservation of energy (as it is done in 
Creutz's version of microcanonical MC~\cite{Creutz1983}).

\section{Results}

We performed microcanonical MC simulations on highly compressed fcc crystals 
whose atoms interact via the Lennard-Jones pair potential truncated at a cutoff
radius $r_c$,

\begin{equation}
\phi(r; \sigma, \epsilon, r_c) =
4\epsilon\Theta(r_c-r)\left[\left(\frac{\sigma}{r}\right)^{12}-\left(\frac{\sigma}{r}\right)^6 \right],
\end{equation}
where $r$ is the distance between atoms $i$ and $j$ and the values considered for the 
parameters are $\sigma$=3.41 \AA~, $\epsilon/k_B$= 119.8 K and $r_c$=2.5 $\sigma$. The 
crystals simulated ranged from 3$\times$3$\times$3 to 6$\times$6$\times$6 unit cells 
(108 to 864 atoms) with a lattice constant $a$=4.2 \AA. This value of $a$
corresponds to a point on the melting curve with 5133 K $< T_m <$ 5251 K and
$P(T_m)=$ 70 GPa, as reported by Belonoshko from two-phase
simulations~\cite{Belonoshko2006}. In both MD and MC simulations we imposed periodic 
boundary conditions. We performed about 20 different simulations under each method and 
system size, with temperatures ranging from 5000K to 7000K. Averages were taken over the 
last 50 thousand steps in each simulation, after 50 thousand equilibration steps. 
For MD simulations, the time step used was $\Delta t$=1 fs. 

As all MC methods based on the Metropolis rule, a reasonably low rejection rate
for moves must be imposed, and the usual way is to employ a small enough atomic
displacement when proposing a move. In our MC simulations, we always kept the rejection
rate below 60\%. Interestingly, we noted that failure to control rejection has a
similar effect to failure of energy conservation in MD, namely, averages like
temperature start to drift linearly with MC ``time''.

\begin{figure}[h]
\vspace{30pt}
\begin{center}
\includegraphics[scale=0.16]{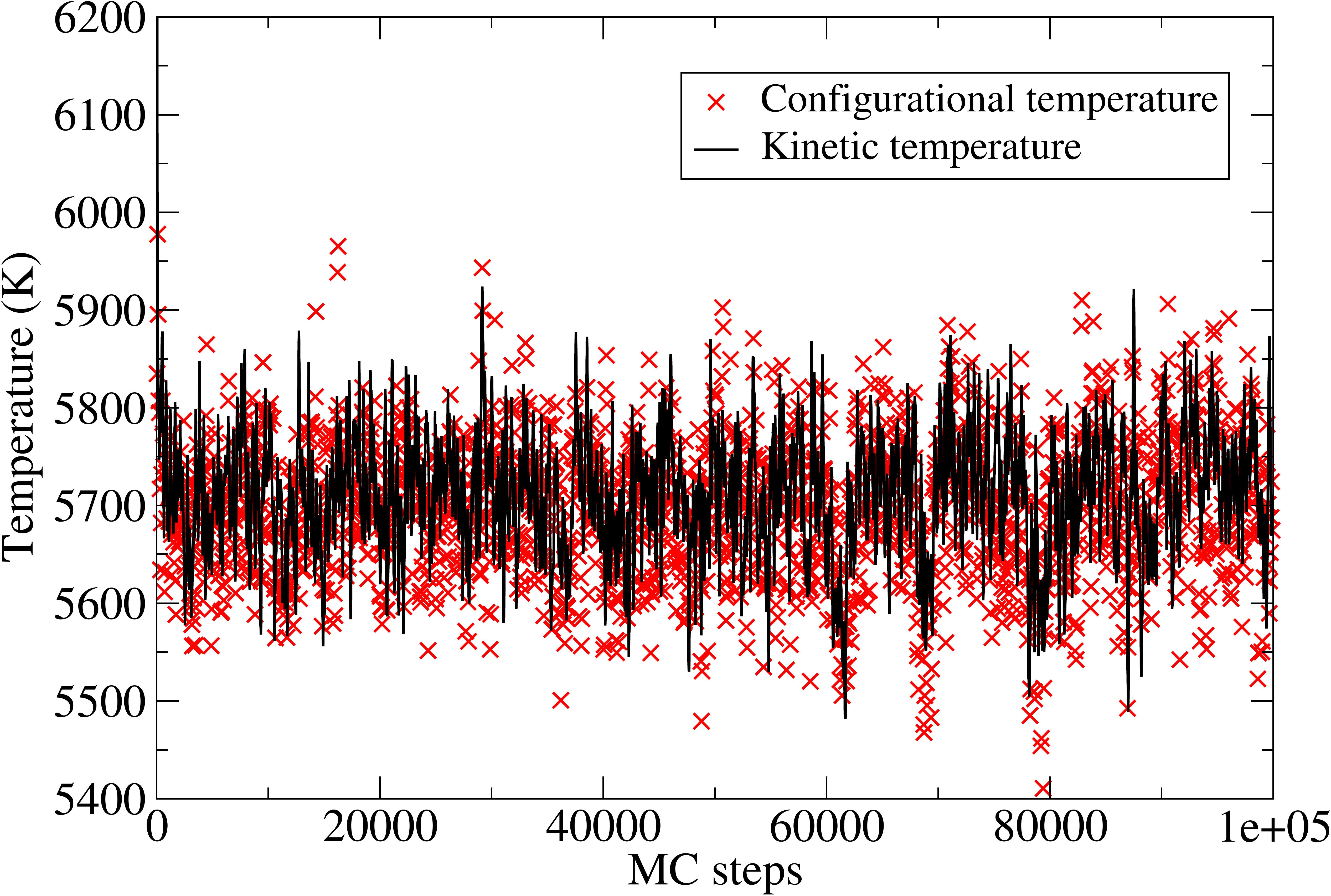}
\end{center}
\caption{Comparison between configurational and kinetic instantaneous
temperatures during a MC simulation for $N$=864 atoms.}
\label{ConfigTemp}
\end{figure}

Instantaneous temperatures $T_i$ during a MC simulation can be obtained from the kinetic
energy (as usual in molecular dynamics simulations), 

\begin{equation}
\frac{1}{k_BT_i}=\frac{3N-2}{2(E-\Phi)},
\end{equation}
but also from derivatives of the potential energy, using the
so-called configurational temperature~\cite{Rugh1997, Butler1998, Rickayzen2001})
which is given by

\begin{equation}
\frac{1}{k_BT_i}=\frac{\nabla^2\Phi}{\vert\vec{\nabla}\Phi\vert^2} +
\mathcal{O}(\frac{1}{N}),
\end{equation}

and from these, the equilibrium thermodynamical temperature is obtained as
microcanonical averages, $T(E)=\big<T_i\big>_E$. Figure \ref{ConfigTemp} shows a
comparison of both configurational and kinetic definitions of instantaneous
temperature for a typical MC run. This provides an additional consistency check
for our MC simulations, in order to make sure the microcanonical ensemble is
adequately sampled.

\begin{figure}[t]
\vspace{25pt}
\begin{center}
\includegraphics[scale=0.14]{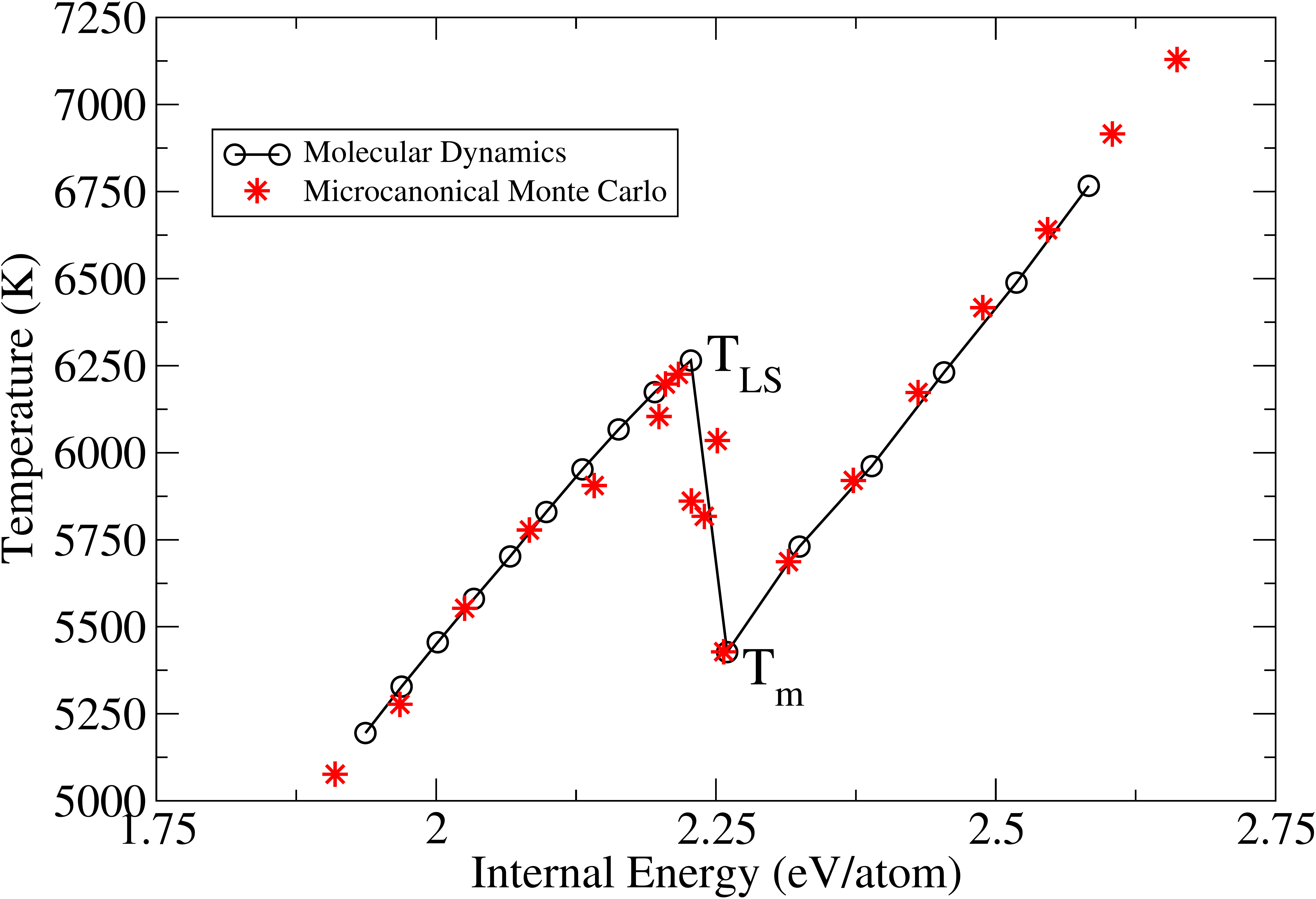}
\end{center}
\caption{Comparison between the isochoric curves obtained by standard,
microcanonical molecular dynamics and microcanonical MC methods, for a
system of $N$=864 atoms.}
\label{ZRay}
\end{figure}

Figure \ref{ZRay} shows a comparison between the isochoric curves obtained by
standard, MD version of the Z method, and the MC version, for a system size $N$=864 atoms. 
The agreement between the two is perfect in the thermodynamic stability region 
(solid and liquid straight lines), and both methods yield the same $T_m$ and
$T_{LS}$ within the statistical margin of error, as shown in table
\ref{tbl_temps}. The same level of agreement is seen for all the smaller system sizes studied.
The slight overestimation (about 4.5\%) of $T_m$ as compared with Belonoshko's two-phase
simulations with $N$=32000 is a well known size effect, the Z method overestimates $T_m$ 
and $T_{LS}$ for small systems. Here we are only interested in comparing the two
implementations of the method for equal conditions.

\begin{table}
\vspace{50pt}
\begin{center}
\begin{tabular}{|c|c|c|}
\hline
Method & $T_{LS}$ (K) & $T_m$ (K) \\
\hline
Molecular Dynamics & 6265 $\pm$ 148 & 5427 $\pm$ 103 \\ 
\hline
Monte Carlo & 6225 $\pm$ 138 & 5428 $\pm$ 123 \\
\hline
\end{tabular}
\end{center}
\caption{Values of the critical superheating temperature $T_{LS}$ and the
melting temperature $T_m$ obtained for $N$=864 atoms by the MD Z method and the
MC Z method.}
\label{tbl_temps}
\end{table}

\begin{figure}[h]
\vspace{30pt}
\begin{center}
\includegraphics[scale=0.16]{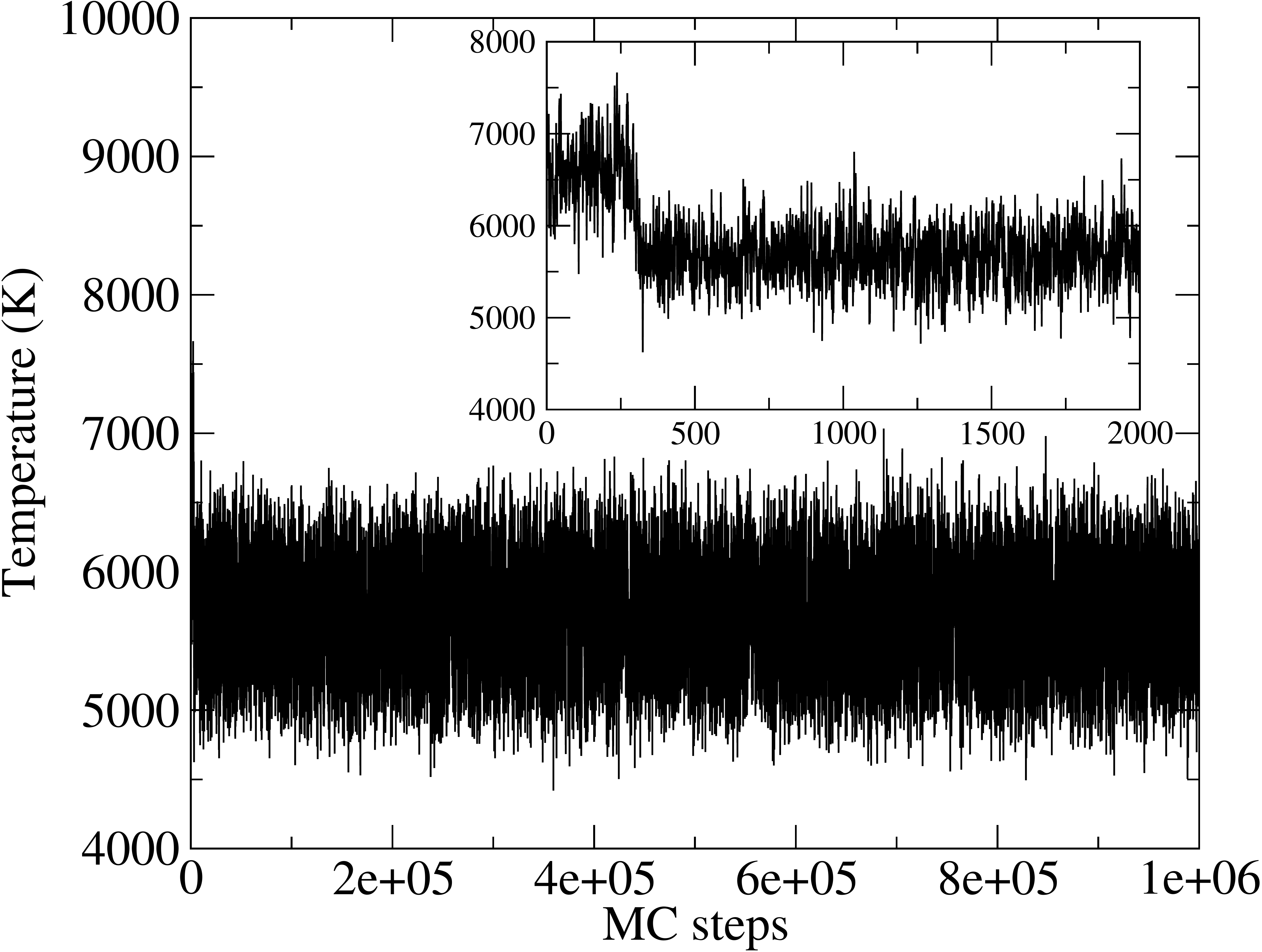}
\end{center}
\caption{Evolution of the instantaneous temperature during a MC
simulation where melting is observed, for $N$=108 atoms. Total simulation comprises
1 million MC steps, and the inset shows the first 2000 MC steps.}
\label{ZFluct}
\end{figure}

\begin{figure}[h]
\vspace{30pt}
\begin{center}
\includegraphics[scale=0.16]{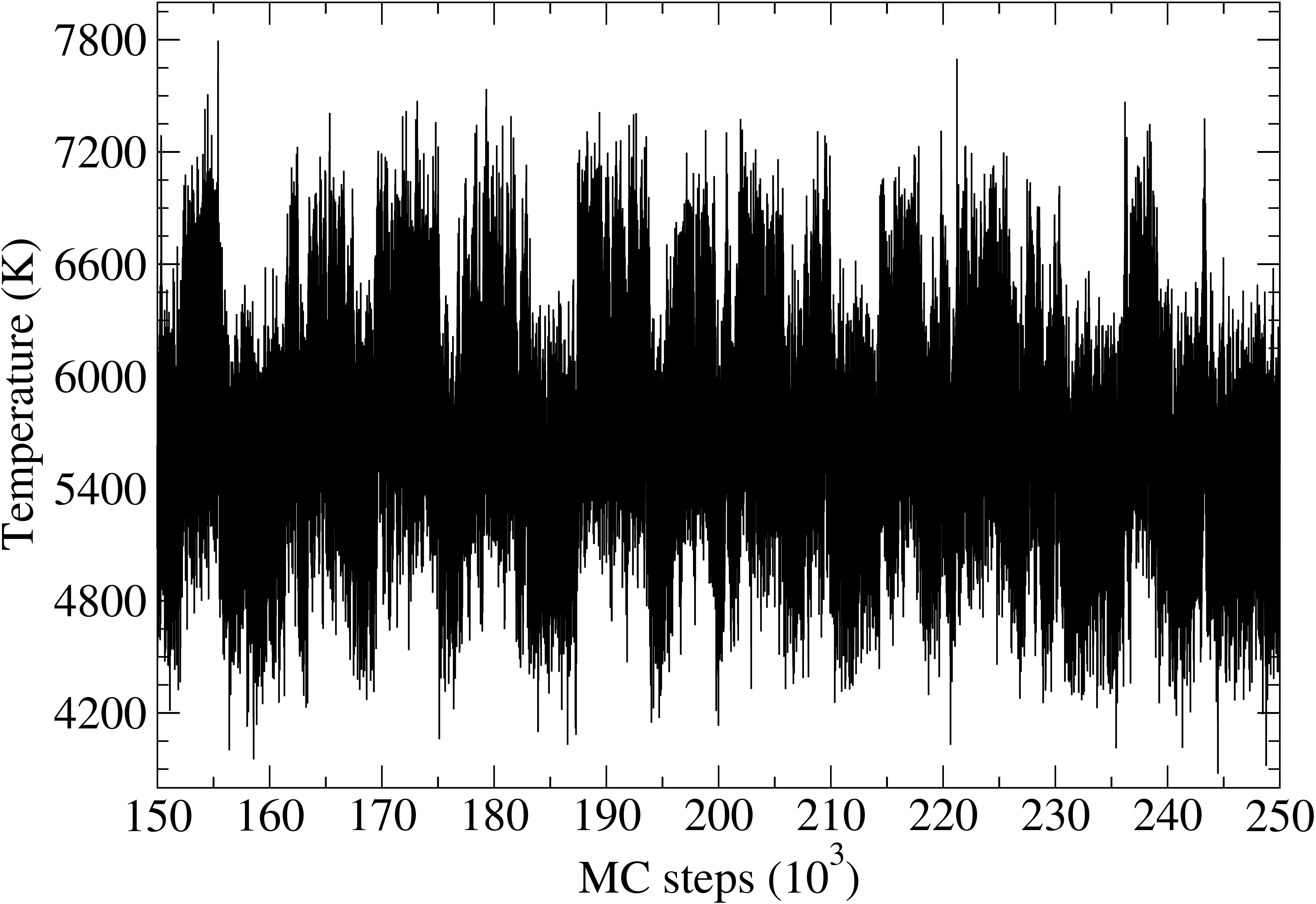}
\end{center}
\caption{Alternation between solid and liquid phases as seen from the
instantaneous temperature, for a MC simulation with $N$=72 atoms.}
\label{fig_altern}
\end{figure}

Figure \ref{ZFluct} shows the evolution of the instantaneous temperature as a
function of MC steps, for a total energy above $E_{LS}$. The system starts in
the solid phase, melting spontaneously after the first 350 steps. Temperature
fluctuations are significant, due to the limited size of the system. In this case we 
did not see the alternating behavior between solid and liquid phases expected in
small systems (as reported by Alf\`{e}~\cite{Alfe2011}), most probably because
of the larger system size simulated (864 atoms instead of 96). In fact, for 72
atoms the alternation occurs, as shown in Fig. \ref{fig_altern}. The precision 
needed to find the energy $E_m$ at which dynamical coexistence is observed
depends on the system size, this is due to the fact that there is a finite, non-zero 
probability that a small system could oscillate between phases even when their 
respective entropies are not exactly equal (i.e. when we are close but not exactly at 
$E_m$). In Ref.~\cite{Alfe2011}, the alternation effect is treated considering
the fraction of time $\alpha$ spent in its solid or liquid phase, and from
equilibrium microcanonical considerations, a relation connecting the fractions
$\alpha$ to the entropy of melting $\Delta S$ is found. Interestingly, the
analysis can be done without any reference to equilibrium, in the framework of
Evans' fluctuation theorem~\cite{Evans2002},

\begin{equation}
P(S\rightarrow L)/P(L\rightarrow S) = e^{2N\Delta s(S\rightarrow L)/k_B}.
\label{eq_evans}
\end{equation}

Here $P$ is the transition probability from one state to another, $L$ and $S$ represent 
liquid and solid states and $\Delta s$ is the entropy of melting per atom. From
Eq. \ref{eq_evans} it can be seen that, for small systems ($N \rightarrow 0$), $\Delta s$ 
can be slightly larger and the right-hand side will still be close enough to 1 to allow 
transitions from solid to liquid and the reverse. In practice, finding this alternation in 
MC simulations could be more difficult also due to our use of simple local updates (one atom is displaced
at a time on each trial move) which near the transition point could lead to an analog 
of the \emph{critical slowing down} effect seen in lattice MC simulations~\cite{Swendsen1987, 
Wolff1989, Binder2010}, thus making alternation events extremely difficult to
generate.

It is important to notice that reproducing $T_{LS}$ via a stochastic procedure
does not contradict the notion of $T_{LS}$ being related to time scales~\cite{Davis2011} 
(which are absent in an absolute sense in Monte Carlo simulations). Far from it,
a correct prediction of $T_{LS}$ under stochastic dynamics seems to support the notion of 
it being related to random walk statistics with jump probabilities only dependent on 
(microcanonical) thermodynamic properties.

In fact, as an illustration consider simulations of the mean square displacement $\big<r^2(t)\big>$ 
(a) in liquids via molecular dynamics, and (b) via isotropic random walk simulations. 
In both cases we have $$\big<r^2(t)\big> \propto 6Dt$$ as $t \rightarrow \infty$, and this is not surprising 
even though $t$ is not a ``real'' time but a number of MC steps. In both cases a
random walk is used to sample a thermodynamical quantity, namely $D=D(T)$, which
happens to have dynamical consequences such as the diffusion rate. In the same way, we conjecture 
that $T_{LS}$ is a function of dynamical properties which in turn, depend only on features of 
the material's potential energy landscape.

\section{Summary and conclusions}

Belonoshko \emph{et al}~\cite{Belonoshko2007} attributed the success of the Z
method in reaching the highest $T_{LS}$ among several molecular dynamics methods
to the preservation of the natural dynamics of the system. However, we have
shown this is not the case: using a MC algorithm which is
completely oblivious to the equations of motion we have reproduced the same
$T_{LS}$ and the same $T_m$ as in standard molecular dynamics. This suggests that 
the previously thought advantage of the Z method comes from a different direction, namely, 
that the preservation of the microcanonical condition is the important fact, and
the specific trajectory followed by each atom is not important. Therefore, any 
method capable of computing microcanonical averages should be just as reliable in Z method computations.

Practical consequences of this finding are clear in terms of the efficiency for 
large or complex systems. The success of MC methods in the determination of
the Z curve hints to the possibility of a fully-parallelizable version of the method.

\section{Acknowledgements}

SD acknowledges financial support from FONDECYT grant 1140514. GG and SD thank partial support from CONICYT-PIA grant
ACT-1115, Chile.

\bibliography{allrefs}
\bibliographystyle{apsrev}

\end{document}